\documentclass[prl,twocolumn,english,showpacs,nofootinbib]{revtex4-2}
\usepackage{amssymb}
\usepackage{color}
\usepackage{graphicx}
\usepackage{bbold}
\usepackage{bm}
\usepackage{amsmath}
\usepackage{amsfonts}
\usepackage{amssymb}
\usepackage{braket}
\usepackage{units}
\usepackage{csquotes}
\usepackage{upgreek}
\usepackage{xcolor}
\usepackage{fancyhdr}
\usepackage{float}

\usepackage[normalem]{ulem}
\usepackage{lipsum}

\begin{document}
	
	
	\title{Machine-learning the phase diagram of a strongly-interacting  Fermi gas}
	
	\author{M. Link$^1$, K. Gao$^{1,2}$, A. Kell$^1$, M. Breyer$^1$, D. Eberz$^1$, B. Rauf$^1$, and M. K\"ohl$^1$}
	\affiliation{Physikalisches Institut, University of Bonn, Wegelerstra{\ss}e 8, 53115 Bonn, Germany}
	\affiliation{Department of Physics, Renmin University of China, Beijing 100872, China}


\begin{abstract}
We determine the phase diagram of strongly correlated fermions in the crossover from Bose-Einstein condensates of molecules (BEC) to Cooper pairs of fermions (BCS) utilizing an artificial neural network. By applying advanced image recognition techniques to the momentum distribution of the fermions, a quantity which has been widely considered as featureless for providing information about the condensed state, we measure the critical temperature and show that it exhibits a maximum on the bosonic side of the crossover. Additionally, we back-analyze the trained neural network and demonstrate that it interprets physically relevant quantities.
\end{abstract}

\maketitle

When an ensemble of attractively interacting fermions is cooled to below a critical temperature $T_c$ it   transitions from a normal phase into a superfluid or superconducting phase. The precise value of the phase transition temperature is governed by the microscopic details of the system, such as the interaction strength and interparticle correlations, and  can exhibit non-trivial dependencies. For example, in the crossover from BCS to BEC, it has been theoretically predicted that the critical temperature depends non-monotonically on the  interaction parameter  \cite{Melo1993,Haussmann2007,Burovski2008,Bulgac2008,Floerchinger2008,Floerchinger2010,Pini2019,Pisani2018}, see Figure 1a. The non-monotonic behaviour is rooted in the fundamental change of the nature of pairing below  the critical temperature. For Cooper pairing (BCS) one expects an exponential dependence of $T_c$ on the interaction strength whereas dimer pairing (BEC) implies a nearly constant $T_c$. The division  between the two regimes is not at unitarity but is expected to be on the BEC  side of the crossover \cite{Son2006,Carlson2008}. In this manuscript we study the critical temperature across the BCS/BEC crossover using an artificial neural network to analyze the momentum-distribution of ultracold atomic Fermi gases.

\begin{figure}
	\includegraphics[width=\columnwidth]{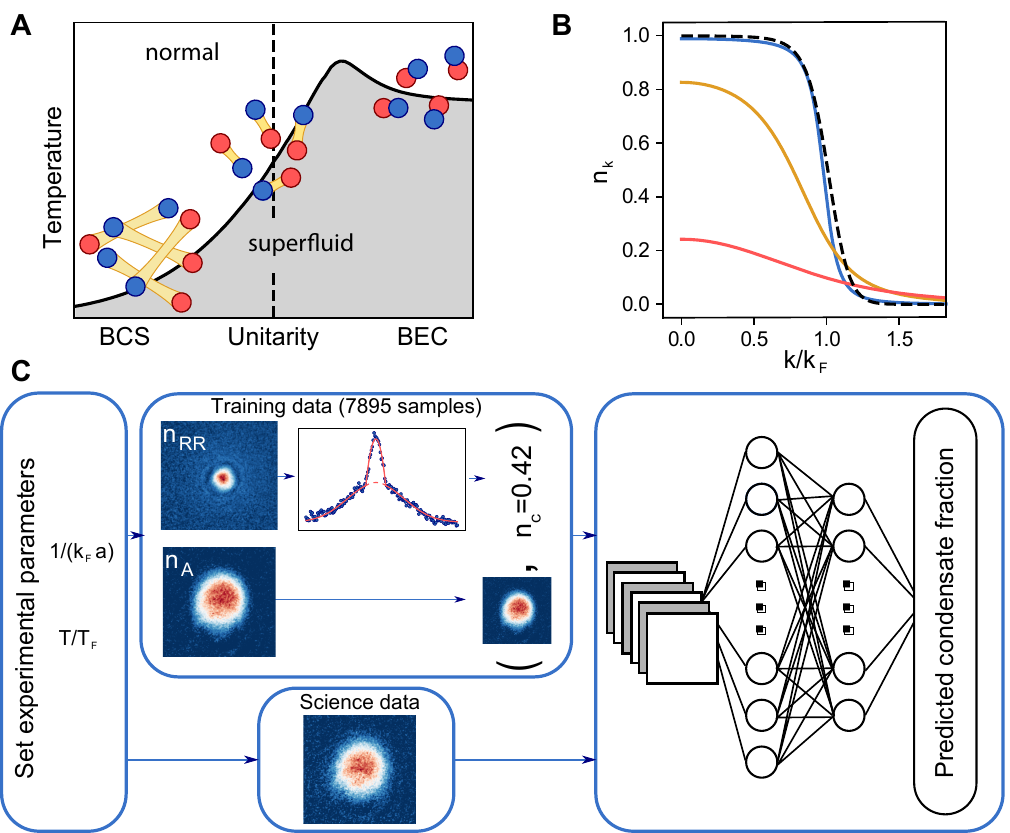}
	\caption{ {\bf A} Sketch of the phase diagram across the BCS/BEC crossover including the critical temperature for condensation (solid line). BCS ground state is dominated by long-range Cooper pairs whereas the BEC exhibits dimer pairing of the fermions.  {\bf B} Momentum distribution of an ideal Fermi gas at a temperature of $T/T_F=0.15$ (dashed line) in comparison with the momentum distribution of the BCS ground state for different interaction strengths: $1/(k_Fa)=-1$ (blue), $1/(k_Fa)=0$ (yellow), and $1/(k_Fa)=1$ (red). {\bf C} Principle of the data analysis using a neural network. Condensate fractions are determined from time-of-flight images after a rapid-ramp. This information is used to label time-of-flight data for equal parameters but without rapid ramp. A neural network is trained on the labeled data to predict the condensate fraction. } 
\end{figure}

A precision determination of the critical temperature across a broad range of interaction strengths have  so far been hindered by insufficient experimental detection capabilities.  One main challenge is that upon release from the trap in a conventional time-of-flight study, Cooper pairs break and are not amenable for direct detection. Nevertheless, they leave a weak imprint onto the momentum distribution of the fermions. In Figure 1b we compare the momentum distribution of a homogeneous Fermi gas at a temperature of $T/T_F=0.15$ (i.e. near the critical temperature) with the momentum distributions of BCS ground state wave functions for different interaction parameters. The pairing signature is by far not as pronounced as the celebrated bimodal momentum distribution of a  Bose-Einstein condensate and therefore the detection of the condensate fraction is much more difficult. Additionally, finite temperature, interactions and the inhomogeneity of the harmonically trapped sample further obscure the pairing signature \cite{Ketterle2008}. In order to detect the minuscule modifications of the momentum distribution in the time-of-flight images, we have developed and applied a neural network for advanced image recognition. We favour neural network processing over standard data fitting since the neural network is unbiased as compared to applying a predetermined fitting function and therefore might detect physical signatures beyond a model-based analysis. Recently,  applications of these sophisticated techniques have entered into the field of quantum physics for the identification of phases of quantum matter \cite{Carleo2017,Tanaka2017,vanNieuwenburg2017,Torlai2018,Wang2016,Huembeli2018,Rem2019}. However, even  when being successfully trained, artificial neural networks have acted as  ``black boxes'' hiding their decision criteria. Specifically, whether or not the network actually identifies physically relevant criteria for computing its output  has remained obscure. Generally, the interpretation of neural networks and their causality is rather challenging and currently a major topic in computer science  \cite{Gilpin2018}. In this work, we demonstrate that the back-analysis of neural networks provides further details of the physics, which are not accessible by conventional means.

Experimentally, we prepare a quantum gas of $\sim 3\cdot 10^5$  atoms per spin state in the two lowest hyperfine states $\ket{1}$ and $\ket{2}$ of $^6$Li in an optical dipole trap, similar to our previous work  \cite{Behrle2018}. We adjust the interaction strength of the sample by Feshbach resonance
and the temperature by changing the trap \cite{Kuhnle2011}. The interaction and temperature are tuned independently of each
other and the thermalized cloud is detected by absorption imaging after
ballistic expansion, see Appendix.

The neural network employed for image analysis comprises of three convolutional and pooling layers and three fully-connected layers  and is trained through stochastic gradient descent with Adam optimizer \cite{adam}, see Appendix. 
In order to train and validate our neural network, we employ a supervised learning method \cite{good16}. To this end, we measure two different density distributions after time-of-flight, see Figure 1c: (1) The density distribution $n_A$ of the atoms directly released from the optical dipole trap. During the expansion, the Cooper pairs are broken and $n_A$ is related to the momentum distribution of the fermions convolved with interaction effects during the expansion. (2) The density distribution $n_{RR}$ after applying the rapid ramp technique \cite{Regal2004,Zwierlein2004,Altman2005,Tikhonenkov2006}, which measures the momentum distribution of the molecules that have been created from the Cooper pairs. Even though this technique is expected to preserve physics in many cases,  quantitatively and principally there are  open questions about the adiabaticity of the ramp and how this might affect weak signatures such as small condensate fractions near the critical temperature. During the training process, we label input pictures of $n_A$ with condensate fractions obtained from bimodal fits to $n_{RR}$ at the same experimental parameters. We exclude data with temperatures near the critical temperature from learning. Moreover, in order to prevent the network from learning unwanted correlations between directly accessible parameters (such as atom number and condensate fraction), we use training data from different interaction values throughout the crossover at $1/(k_Fa)= \{1.6, 1.0, 0.5, 0.0, -0.5, -0.6\}$ 
on a total of 7895 labeled examples. Here, $k_F$ denotes the Fermi wave vector calculated from the atom number and the trap parameters  and $a$ the s-wave scattering length. We extract the critical temperature from the neural network predictions for the direct-release time-of-flight images across the whole range of the BCS/BEC crossover by taking a piecewise linear fit of the condensate fraction5. 

 Qualitatively, the behaviour of the critical temperature of the superfluid transition across the BCS/BEC crossover can be understood by starting from the extreme regimes: in the weakly-attractive BCS limit, the critical temperature scales $k_BT_c \sim E_F \exp[-\pi/(2k_F |a|)]$  \cite{Melo1993,Gorkov1961}. Here, $E_F$ denotes the Fermi energy. In the opposite regime, far on the BEC side, we encounter a weakly-repulsively interacting gas of bosons. The bosons have twice the mass of the fermions $M_B=2 m$ and half the density $n_B=n/2$. The critical temperature of the ideal Bose gas is simply given by $k_BT_c^0 \sim \frac{\hbar^2 n_B^{2/3}}{M_B}$. Unlike in the BCS regime, the critical temperature of the Bose gas has a very weak dependence on the interaction strength between the bosons $T_c(a_B)= T_c^0\left[1+c n_B^{1/3} a_B+ ...\right]$, where $a_B=0.6 \,a$ \cite{Petrov2004} denotes the s-wave scattering length between two bosons, and $c$ is a positive constant \cite{Arnold2001,Kashurnikov2001}. From this simple argument, we expect an increase of the critical temperature when approaching the crossover from the BEC side and hence a maximum critical temperature somewhere in the crossover regime.

From the previous consideration it is obvious that a careful determination of both density and temperature is very important. In the trapped gas of our experiment, the two quantities are inversely related to each other and, furthermore, also interparticle interactions change the density. 

The calibration of density and temperature proceeds in the following way: We take \textit{in-situ} absorption images of the trapped gas along two orthogonal spatial directions (in order to account for asymmetries of the trapped cloud) for different interaction strengths and temperatures. On these data, we perform an inverse Abel transform to reconstruct the  density distribution inside the trap. This serves two purposes: on the one hand, we obtain the center density which we use for the normalisation of the data and on the other hand, the density distribution $n_{\sigma}(r)$ feeds into the temperature calibration in the next step. Then,  we use the data from the unitary Fermi gas [$1/(k_Fa)=0$] and its both theoretically \cite{Haussmann2007,Goulko2010,Burovski2006,Pisani2018} and experimentally \cite{luo2009,horikoshi2010,Nascimbene2010,Ku2012} well known critical temperature of $T_c=0.167\,T_F$  to precisely reconstruct our trapping potential. To this end, the inverse equation of state of the unitary Fermi gas \cite{Ku2012} is applied to the in-trap density distribution reconstructed from \textit{in-situ} high-intensity absorption images of the cloud at $T=T_c$.  In the final step, we use the obtained knowledge of trap potential and  measured \textit{in-situ} density profiles $n_{\sigma}(r)$ to determine the temperature by fitting a virial expansion of the equation of state to the outermost regions of the trapped cloud where the gas is not condensed. 

In Figure 2, we show the results of the critical temperature for a homogeneous gas in comparison with theoretical predictions as a function of the interaction parameter $1/(k_Fa)$. Since the condensation will initiate at regions of highest  density, i.e., at the center of the trapping potential, we adopt a local density approximation and use the density and interaction parameters at the center of the cloud to compare with the theory of the uniform gas. Our results show a steady increase of the critical temperature from the BCS side up to interaction strengths of approximately $1/(k_Fa)=0.5$. There, $T_c$ levels off and stays approximately constant or, possibly, declines weakly for higher coupling strengths. 
Overall, our results are in very good agreement with several theory predictions in different ranges of the phase diagram. Throughout the whole crossover, the agreement with the extended Gorkov-Melik-Bakhudarov (GMB) theory \cite{Pisani2018} is striking and both position and value of the maximum $T_c$ are well compatible with the theoretical results. On the BCS side our data are higher than the  Quantum-Monte Carlo calculations \cite{Burovski2006,Burovski2008,Bulgac2008} and the extended GMB theory, however, close to the theoretical prediction of reference \cite{Haussmann2007}. The experimental results agree with trends observed also in earlier measurements of the phase diagram in both potassium \cite{Regal2004} and Lithium gases \cite{Zwierlein2004} in which the critical temperature on the BCS side did not fall off as rapidly with decreasing interaction strength. Possible suggested explanations include effects of the harmonic trap \cite{Perali2004}, the formation of dimers above the resonance \cite{Zwierlein2004} and non-adiabaticities of the rapid ramp in the training data.

In order to verify our methodology, we have double-checked the performance of the trained neural network using a different Feshbach resonance in another spin mixture of the Lithium atoms: We prepare a condensate in the hyperfine states $\ket{1}$ and $\ket{3}$ for which the position and width of the Feshbach resonance are different and atom number and starting temperature as compared to the training cases are also different. Nevertheless, the neural network successfully predicts the condensate fraction at $1/k_F a = -0.41$ ($k_F$ calculated from the atom number) with the same critical temperature.

\begin{figure}
	\centering
	\includegraphics[width=\columnwidth,clip=true]{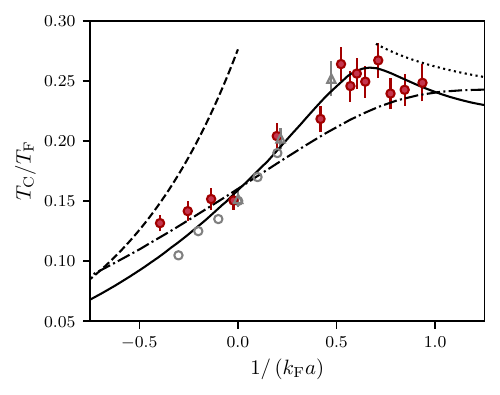}
	\caption{Critical temperature across the BEC-BCS crossover referenced to the homogeneous gas. Full symbols: experimental data. Error bars are calculated from the standard error and an estimation of the systematic error caused by non-harmonicities of the trap. The latter effect dominates and is discussed in more detail in the appendix. Dashed line: BCS theory with GMB corrections; solid line: extended GMB theory \cite{Pisani2018}; dash-dot line: theory from \cite{Haussmann2007}; dotted line: interacting BEC; open triangles: quantum Monte-Carlo data \cite{Burovski2008}, open circles: quantum Monte-Carlo data \cite{Bulgac2008}. }
	\label{Fig3}
\end{figure}

An important remaining question is  whether  the optimized neural network has learned physically relevant quantities. In other words: does the neural network spot hidden details in the data during its optimization and can we extract these information to draw conclusions for the physics? In previous applications of neural networks to analyze quantum problems this had often  not been considered, and, generally, the question of causality in machine learning is becoming increasingly important also in computer science.  We extract from the neural network which neurons have been mostly activated. To this end, we employ a backpropagation-based approach (DeepLIFT, \cite{deeplift}) that assigns importance scores to the inputs for a given output. The importance scores can then be identified to reveal those neurons (or, simply put, regions of the image) that contribute most decisively to the neural network output. In Figure 3, we show the importance scores obtained for different momenta. The results highlight that in the BCS regime [$1/(k_Fa)<0$] the neural network output is dominated by the momentum density near $k\simeq 0.2\,k_F$. In contrast, on the bosonic side  [$1/(k_Fa)>0$], the neural network output mostly relies on low-$k$ data. This finding is in agreement with the expectation of the effects of pairing in the fermionic and bosonic sides of the crossover (see Figure 1b) and indicates that the neural network optimizes indeed for physically relevant features in the time-of-flight data.

\begin{figure}
	\centering
	\includegraphics[width=\columnwidth,clip=true]{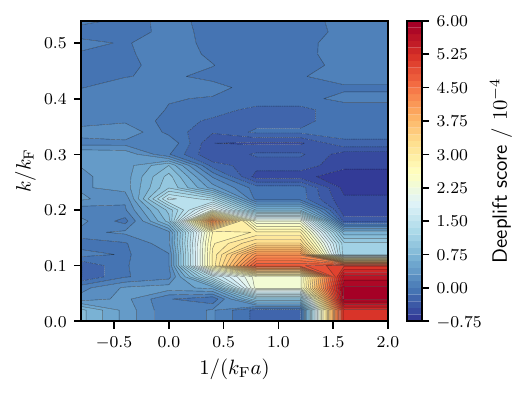}
	\caption{Back-analysis of the neural network. For every interaction strength, we determine which neurons activate most for determining the condensate fraction. On the BEC side the most active neurons are at low momenta whereas at unitarity and on the BCS side the most active are at large momenta.}
	\label{Fig4}
\end{figure}

In summary, we have demonstrated that a neural network can be utilized for the detection of quantum phases of strongly-interacting systems with high accuracy. Moreover, we show that back-analysis of the trained neural network allows to extract insights much beyond standard fitting routines and therefore opens a new route to the precision analysis of physical data.

This work has been supported by the Alexander-von-Humboldt Stiftung, DFG (SFB/TR 185 project C6), Cluster of Excellence Matter and Light for Quantum Computing (ML4Q) EXC 2004/1 – 390534769. We thank Kilian Kluge for discussions.

\section*{Appendix}

\subsection*{Preparation and detection the sample}

The Lithium atoms are confined in a trap formed by two intersecting laser beams of $1070\,\text{nm}$ wavelength with final trap frequencies in harmonic approximation of $2\pi \times (168, 166, 238)\,\text{Hz}$. The final temperature of the sample after evaporation close to a Feshbach resonance at $834\,\text{G}$ is $T/T_F = 0.08 \pm 0.01$. After preparation of the condensate in the crossover regime, we choose a desired interaction strength $1/k_F a$ by adiabatically ramping the magnetic field to the corresponding field value $B_\text{final}$. To controllably increase the temperature, we suddenly decompress and subsequently recompress the trap for a variable amount of time \cite{Kinast2005,Kuhnle2011}. This is followed by a hold time of $50\,\text{ms}$ for thermalisation. We perform detection of the gas by time-of-flight imaging. To this end, the optical dipole trap is extinguished rapidly and the gas expands. Owing to a residual inhomogeneity of our magnetic bias field, the gas expands into a weak harmonic trapping potential with frequencies $2 \pi \times (16, 16, 20i)\,\text{Hz}$.

In order to image the bimodal distribution $n_\text{RR}$, the cloud is subjected to a rapid projection onto Feshbach molecules by suddenly ramping the magnetic field to the zero-crossing of the scattering length at a magnetic field of 534\,G  before $15\,\text{ms}$ time-of-flight. The cloud is then imaged close to the resonance on the repulsive side at a magnetic field strength of 760\,G. To image the density distribution of the atoms after release from the trap $n_\text{A}$, we perform standard time-of-flight imaging after $5\,\text{ms}$ without changing the magnetic field.

For the reconstruction of the equation of state, we perform high-intensity absorption imaging \textit{in-situ} to resolve the very high densities in the trap  \cite{Reinaudi2007}. Because the size of the cloud has to be known in all spatial directions, the high-intensity absorption imaging is done along two perpendicular lines of sight.

\subsection*{The neural network}

\begin{table}
	\centering
	\begin{tabular}{|c|c|c|}
		\hline
		Layer (type) & Output Shape & Parameters\\ \hline
		Input & (Batch, 150, 170, 1) & 0 \\ \hline
		2D Convolutional & (Batch, 150, 170, 30) & 300\\\hline
		Max Pooling & (Batch, 75, 85, 30) & 0 \\\hline
		2D Convolutional & (Batch, 75, 85, 40) & 30040\\\hline
		Max Pooling & (Batch, 37, 42, 40) & 0 \\\hline
		2D Convolutional & (Batch, 37, 42, 50) & 50050 \\\hline
		Max Pooling & (Batch, 18, 21, 50) & 0 \\\hline
		Flatten & (Batch, 18900) & 0 \\\hline
		Dense & (Batch, 600) & 11340600 \\\hline
		Dropout & (Batch, 600) & 0 \\\hline
		Dense & (Batch, 300) & 180300 \\\hline
		Dropout & (Batch, 300) & 0 \\\hline
		Dense & (Batch, 1) & 301 \\\hline
	\end{tabular}
	\caption[Neural network architecture used for the phase diagrams]{\textbf{Neural network architecture used for the phase diagrams.} \label{tab:architecture}}
\end{table}

The neural network architecture used to generate the phase diagram is given in table \ref{tab:architecture} and is realised with the TensorFlow library \cite{tensorflow}. It consists of three convolutional layers combined with max pooling layers followed by two dense layers with dropout regularisation and one output neuron. The total number of tunable parameters is $11601591$. We train the network with stochastic gradient descent using Adam optimizer with learning rate $1.5\cdot 10^{-4}$ \cite{adam} on the mean squared error loss function. We use 7895 labelled datapoints in total from which $90\,\%$ are used for training and $10\,\%$ for validation. We train each network for 15 epochs with a batch size of 30. The training data is shuffled after each epoch. We tested several network architectures and generally found the performance robust against changes of the hyper parameters. We conclude that network architectures similar to the one used here provide robust learners for the detection of the condensate fraction. Moreover, we have taken data sets with different magnifications of the optical imaging system. The images were then scaled prior to feeding the data into the neural network and transfer learning on 2500 images was performed, while freezing the parameters of the convolutional layers. We found our data to be robust against this scaling operation.

\subsection*{Equation of state reconstruction}


We reconstruct the equation of state under local density approximation $n_\sigma(\mu-V)$ for all temperatures and interaction strengths entering the phase diagram. Here, $n_\sigma$ is the density distribution of the cloud in the trap, $\mu$ the chemical potential, and $V$ the external trapping potential. From the equation of state, we can extract the temperature of the cloud $T$ and the density in the centre of the trap $n_\sigma|_{V=0}$, which are the quantities needed to calibrate the phase diagram. The Fermi wave vector $k_F$ and Fermi temperature $T_F$ are related to $n_\sigma|_{V=0}$ via $k_F=(3\pi^2n_\sigma|_{V\!=\!0})^{1/3}$ and $T_F=\frac{\hbar^2k_F^2}{2mk_B}$ where $\hbar$ denotes the reduced Planck constant, $k_B$ the Boltzmann constant, and $m$ the mass of $^6$Li atoms.

To determine the temperature of the cloud we utilize that, close to the surface of the cloud, the equation of state can be approximated by a virial expansion $n_\sigma\lambda^3=\sum_n{n b_n z^n}$ with $\lambda=\sqrt{\frac{2\pi\hbar^2}{m k_B T}}$ the thermal de-Broglie wavelength, $b_n$ the $n$-th virial coefficient, and $z=\exp((\mu-V)/k_BT)$. By fitting the virial expansion up to order $n$=3 (or $n$=4 for clouds furthest in the BCS regime) to the dependence of $n_\sigma$ on $V$ close to the surface of the cloud we get the temperature of the cloud $T$. While $b_1$=1 for all scattering lengths $a$ and temperatures $T$, higher order $b_n$s depend on $a$ and $T$. For $b_2$ an analytic formula is known \cite{Lee2006}, $b_3$ has been calculated for a wide range of $a$ and $T$ \cite{Leyronas2011} and also for $a<0$ in \cite{Hou2020} where $b_4$ for $a<0$ is calculated as well.

To perform the equation of state reconstruction, intrap density and potential have to be known. The density is determined from \textit{in-situ} optical density images by first taking the elliptic radial average - respecting the cloud's aspect ratio - and converting the optical density to column density $n_\mathrm{col}(r)$, i.e. the density integrated along the camera's line of sight. The actual density $n_\sigma(r)$ is then reconstructed from $n_\mathrm{col}(r)$ by an elliptic inverse Abel transform
\begin{equation}
	n_\sigma(r)=-\frac{\sigma_r}{\sigma_z}\frac{1}{\pi}\int_r^{\infty}\!\mathrm{d}r' \frac{1}{\sqrt{r'^2-r^2}}\frac{\partial n_\mathrm{col}(r')}{\partial r'}
	\label{eq:inverseAbel}
\end{equation}
where $\sigma_r\,(\sigma_z)$ is the size of the cloud in radial direction (the camera's line of sight).\\
The external potential is only known in harmonic approximation from the trap frequencies, but this approximation is not valid for the crossed Gaussian-beam dipole trap in the region where we perform our thermometry. We therefore derive the full external trapping potential $V(r)$ from the density $n_{\sigma}(r)$ of a cloud with $1/k_Fa=0$ and $T=T_c$. Since the critical temperature of a homogeneous Fermi gas at unitarity is well known to be $T_c=0.167\;T_F$ \cite{Ku2012}, it is sufficient to determine $T_F$ (resp. $n_\sigma|_{V\!=\!0}$) instead of $T$. From reference \cite{Ku2012} the chemical potential of the unitary Fermi gas at $T_c$ can also be related to $T_F$ by $\mu|_{T=T_c}=0.416\;k_BT_F$. With known $\mu$ and $T$, combining the local density approximation $\mu\rightarrow\mu-V(r)$ with the known equation of state of the unitary Fermi gas \cite{Ku2012} $n_{\sigma}(\mu,T)=n_{\sigma}(\mu/k_BT)\rightarrow n_{\sigma}(\,(\mu-V(r))/k_BT\,)$ yields a relation between $n_{\sigma}(r)$ and $V(r)$ without free parameters. Inversion of this relation allows the derivation of $V(r)$ from $n_{\sigma}(r)$ of a cloud with $1/k_Fa=0$ and $T=T_c$.\\
Because $V(r)$ is identical for all clouds, it is possible to fit the equation of state's virial expansion close to the surface of the cloud, where $\exp((\mu-V(r))/k_BT)\ll1$ ensures the validity of the virial expansion, with the free parameters $\mu$ and $T$. $T_F$ can again be calculated from the density at $V=0$ to get $T/T_F$ for arbitrary interaction strengths and temperatures.



It should be noted that the inverse Abel transform assumes elliptic equipotential lines, a condition only approximately fulfilled in our crossed Gaussian-beam dipole trap. We estimate the influence of this systematic error by simulating the column density of a cloud with $1/k_Fa=0$ and $T=T_c$ in a trap comparable to the one used in the experiment. We then use the same procedure as for the experimental data to reconstruct the external potential and perform our thermometry on simulated column densities of an ideal Fermi gas. The error in T introduced in this way stays below 5\%. We therefore use this value as an upper bound to estimate the systematic error of our thermometry. It is the dominating contribution to the uncertainty of the critical temperature as depicted in fig. 2.

The inverse Abel transform also relies on the differential of the column density $\partial n_\mathrm{col}/\partial r$ which is very susceptible to noise when extracting it from the experimental data. We therefore average ca. 30 identically prepared clouds and perform radial averaging. However, experimental noise still dominates close to the center of the cloud where radial averaging has little effect. To determine the central density more reliably, we linearly extrapolate the measured $n_{\sigma}(V)$ data towards $V=0$.


\begin{thebibliography}{10}
	
	\bibitem{Melo1993}
	C.~A.~R. S\'a~de Melo, M.~Randeria, J.~R. Engelbrecht, {\it Phys. Rev. Lett.\/}
	{\bf 71}, 3202 (1993).
	
	\bibitem{Haussmann2007}
	R.~Haussmann, W.~Rantner, S.~Cerrito, W.~Zwerger, {\it Phys. Rev. A\/} {\bf
		75}, 023610 (2007).
	
	\bibitem{Burovski2008}
	E.~Burovski, E.~Kozik, N.~Prokof'ev, B.~Svistunov, M.~Troyer, {\it Phys. Rev.
		Lett.\/} {\bf 101}, 090402 (2008).
	
	\bibitem{Bulgac2008}
	A.~Bulgac, J.~E. Drut, P.~Magierski, {\it Phys. Rev. A\/} {\bf 78}, 023625
	(2008).
	
	\bibitem{Floerchinger2008}
	S.~Floerchinger, M.~Scherer, S.~Diehl, C.~Wetterich, {\it Phys. Rev. B\/} {\bf
		78}, 174528 (2008).
	
	\bibitem{Floerchinger2010}
	S.~Floerchinger, M.~M. Scherer, C.~Wetterich, {\it Phys. Rev. A\/} {\bf 81},
	063619 (2010).
	
	\bibitem{Pini2019}
	M.~Pini, P.~Pieri, G.~C. Strinati, {\it Phys. Rev. B\/} {\bf 99}, 094502
	(2019).
	
	\bibitem{Pisani2018}
	L.~Pisani, A.~Perali, P.~Pieri, G.~C. Strinati, {\it Phys. Rev. B\/} {\bf 97},
	014528 (2018).
	
	\bibitem{Son2006}
	D.~T. Son, M.~A. Stephanov, {\it Phys. Rev. A\/} {\bf 74}, 013614 (2006).
	
	\bibitem{Carlson2008}
	J.~Carlson, S.~Reddy, {\it Phys. Rev. Lett.\/} {\bf 100}, 150403 (2008).
	
	\bibitem{Ketterle2008}
	W.~Ketterle, M.~W. Zwierlein, {\it Proceedings of the International School of
		Physics "Enrico Fermi"\/} (2008), vol. 164.
	
	\bibitem{Carleo2017}
	G.~Carleo, M.~Troyer, {\it Science\/} {\bf 355}, 602 (2017).
	
	\bibitem{Tanaka2017}
	A.~Tanaka, A.~Tomiya, {\it Journal of the Physical Society of Japan\/} {\bf
		86}, 063001 (2017).
	
	\bibitem{vanNieuwenburg2017}
	E.~van Nieuwenburg, Y.-H. Liu, S.~Huber, {\it Nature Physics\/} {\bf 13}, 435
	(2017).
	
	\bibitem{Torlai2018}
	G.~Torlai, {\it et~al.\/}, {\it Nature Physics\/} {\bf 14}, 447–450 (2018).
	
	\bibitem{Wang2016}
	L.~Wang, {\it Phys. Rev. B\/} {\bf 94}, 195105 (2016).
	
	\bibitem{Huembeli2018}
	P.~Huembeli, A.~Dauphin, P.~Wittek, {\it Phys. Rev. B\/} {\bf 97}, 134109
	(2018).
	
	\bibitem{Rem2019}
	B.~S. Rem, {\it et~al.\/}, {\it Nature Physics\/} {\bf 15}, 917 (2019).
	
	\bibitem{Gilpin2018}
	L.~H. Gilpin, {\it et~al.\/}, Explaining explanations: An overview of
	interpretability of machine learning (2018).
	
	\bibitem{Behrle2018}
	A.~Behrle, {\it et~al.\/}, {\it Nature Physics\/} {\bf 14}, 781 (2018).
	
	\bibitem{Kuhnle2011}
	E.~D. Kuhnle, {\it et~al.\/}, {\it Phys. Rev. Lett.\/} {\bf 106}, 170402
	(2011).
	
	\bibitem{adam}
	D.~P. Kingma, J.~Ba, Adam: A method for stochastic optimization (2017), arXiv: 1412.6980.
	
	\bibitem{good16}
	I.~Goodfellow, Y.~Bengio, A.~Courville, {\it Deep Learning\/} (MIT Press,
	2016).
	
	\bibitem{Regal2004}
	C.~A. Regal, M.~Greiner, D.~S. Jin, {\it Phys. Rev. Lett.\/} {\bf 92}, 040403
	(2004).
	
	\bibitem{Zwierlein2004}
	M.~Zwierlein, {\it et~al.\/}, {\it Phys. Rev. Lett.\/} {\bf 92}, 120403 (2004).
	
	\bibitem{Altman2005}
	E.~Altman, A.~Vishwanath, {\it Phys. Rev. Lett.\/} {\bf 95}, 110404 (2005).
	
	\bibitem{Tikhonenkov2006}
	I.~Tikhonenkov, E.~Pazy, Y.~B. Band, M.~Fleischhauer, A.~Vardi, {\it Phys. Rev.
		A\/} {\bf 73}, 043605 (2006).
	
	\bibitem{Gorkov1961}
	L.~P. Gorkov, T.~K. Melik-Barkhudarov, {\it Sov. Phys. JETP\/} {\bf 13}, 1018
	(1961).
	
	\bibitem{Petrov2004}
	D.~S. Petrov, C.~Salomon, G.~V. Shlyapnikov, {\it Phys. Rev. Lett.\/} {\bf 93},
	090404 (2004).
	
	\bibitem{Arnold2001}
	P.~Arnold, G.~Moore, {\it Phys. Rev. Lett.\/} {\bf 87}, 120401 (2001).
	
	\bibitem{Kashurnikov2001}
	V.~A. Kashurnikov, N.~V. Prokof'ev, B.~V. Svistunov, {\it Phys. Rev. Lett.\/}
	{\bf 87}, 120402 (2001).
	
	\bibitem{luo2009}
	L.~Luo, J.~E. Thomas, {\it Journal of Low Temperature Physics\/} {\bf 154}, 1
	(2009).
	
	\bibitem{horikoshi2010}
	M.~Horikoshi, S.~Nakajima, M.~Ueda, T.~Mukaiyama, {\it Science\/} {\bf 327},
	442 (2010).
	
	\bibitem{Nascimbene2010}
	S.~Nascimbene, N.~Navon, K.~Jiang, F.~Chevy, C.~Salomon, {\it Nature\/} {\bf
		463}, 1057 (2010).
	
	\bibitem{Ku2012}
	M.~J.~H. Ku, A.~T. Sommer, L.~W. Cheuk, M.~W. Zwierlein, {\it Science\/} {\bf
		335}, 563 (2012).
	
	\bibitem{Goulko2010}
	O.~Goulko, M.~Wingate, {\it Phys. Rev. A\/} {\bf 82}, 053621 (2010).
	
	\bibitem{Burovski2006}
	E.~Burovski, N.~Prokof'ev, B.~Svistunov, M.~Troyer, {\it Phys. Rev. Lett.\/}
	{\bf 96}, 160402 (2006).
	
	\bibitem{Pilati2008}
	S.~Pilati, S.~Giorgini, N.~Prokof'ev, {\it Phys. Rev. Lett.\/} {\bf 100},
	140405 (2008).
	
	\bibitem{Perali2004}
	A.~Perali, P.~Pieri, L.~Pisani, G.~C. Strinati, {\it Phys. Rev. Lett.\/} {\bf
		92}, 220404 (2004).
	
	\bibitem{Baym1999}
	G.~Baym, J.-P. Blaizot, M.~Holzmann, F.~Lalo\"e, D.~Vautherin, {\it Phys. Rev.
		Lett.\/} {\bf 83}, 1703 (1999).
	
	\bibitem{deeplift}
	A.~Shrikumar, P.~Greenside, A.~Kundaje, Learning important features through
	propagating activation differences, arXiv: 1704.02685 (2017).
	
	\bibitem{Kinast2005}
	J.~Kinast, {\it et~al.\/}, {\it Science\/} {\bf 307}, 1296 (2005).
	
	\bibitem{Reinaudi2007}
	G.~Reinaudi, T.~Lahaye, Z.~Wang, and D. Gu$\acute{\text e}$ry-Odelin, Strong absorption imaging of dense clouds of ultracold atoms", {\it Opt. Lett.\/} {\bf 32}, 3143-3145 (2007)
	
	\bibitem{tensorflow}
	M.~Abadi, {\it et~al.\/}, {TensorFlow}: Large-scale machine learning on
	heterogeneous systems (2015). Software available from tensorflow.org.
	
	\bibitem{Lee2006}
	D.~Lee, T.~Sch\"afer, {\it Phys. Rev. C\/} {\bf 73}, 015201 (2006)
	
	\bibitem{Leyronas2011}
	X.~Leyronas, {\it Phys. Rev. A\/} {\bf 84}, 053633 (2011)
	
	\bibitem{Hou2020}
	Y.~Hou, J.~E.~Drut, {\it Phys. Rev. Lett.\/} {\bf 125}, 050403 (2020)
	
		
\end{thebibliography}
\end{document}